\documentclass[american,PRE,twocolumn]{revtex4-2}
\usepackage[T1]{fontenc}
\usepackage[utf8]{inputenc}
\usepackage{amstext}
\usepackage{graphicx}
\usepackage{esint}

\makeatletter
\usepackage{braket}

\makeatother

\usepackage{babel}
\begin{document}
\title{A master equation approach to the $n-$coalescent problem}
\author{Bahram Houchmandzadeh}
\address{Univ. Grenoble Alpes, CNRS, LIPHY, F-38000 Grenoble, France}
\begin{abstract}
Given an evolutionary model, such as Wright--Fisher (WF) or Moran,
the $n-$coalescent problem consists of going backward in time to
find, for example, the time to the most recent common ancestor (MRCA)
and the topology of the tree. In the literature, this problem is 
mainly addressed by directly computing the random variable $t$, the time to reach
the MRCA. I show here that by shifting the focus from the random variable
$t$ to the joined variable $(n,t)$, where $n$ is the number of
ancestors at time $t$, the problem is greatly simplified. Indeed,
$P(n,t)$, the probability of this variable, obeys a simpler master
equation that can be solved in a straightforward way for the most
general model. This probability can then be used to compute relevant
information of the $n-$coalescent, for both random variables $t_{n}$
(random time to reach a given state $n$) and $n_{t}$ (random number
of ancestors at a given time $t$). The cumulative distribution function
for $t_{1}$ for example is $P(1,t)$. In this article I give the 
general solution for continuous time models such as Moran and discrete
time ones such as WF. 
\end{abstract}
\maketitle

\section{Introduction.}

\begin{figure}
\begin{centering}
\includegraphics[width=1\columnwidth]{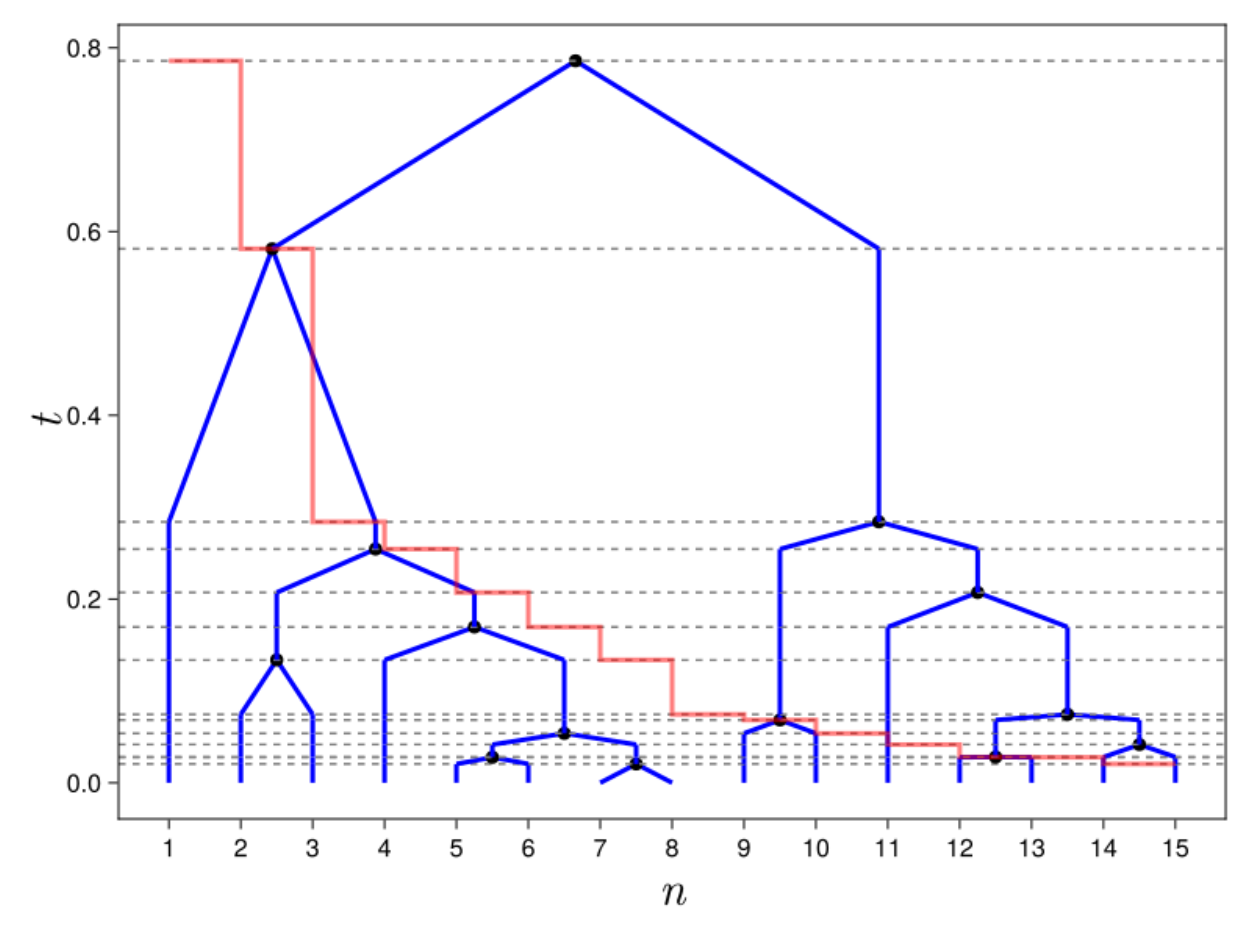}
\par\end{centering}
\caption{Illustration of a Moran coalescent tree (blue lines) with initial
condition $n_{0}=15$ and population size $N=25$. Coalescent events
are marked by black circles and the corresponding coalescent times
by dotted horizontal lines. The red line $n(t)$ is a random path
corresponding to the number of ancestors as a function of (backward)
times for this particular realization.}\label{fig:Illustration-tree}
\end{figure}

The $n-$coalescent problem came to the forefront of population genetics
by Kingman's article\citep{kingman1982} nearly forty years ago
(figure \ref{fig:Illustration-tree}). The framework is now widely
used to capture the stochastic genealogy of $n$ samples drawn at
random from a population of $N$ individuals\citep{fu1999,ewens2004,tavare2004,hein2005,wakeley2009,nordborg2019}.
From a mathematical point of view, the framework can be simply stated:
a given evolutionary model, such as Wright--Fisher or Moran's,
specifies the jump probabilities $W_{n}^{p}$ of having $p$ ancestor
at a prior time $t-\tau$ (either continuous or discrete), given that
the number of individuals at time $t$ is $n$. Using these jump probabilities, one could then compute, through a series of coalescent events, the
distribution of the random variable $t_{n_{0},1}$ of (backward) time
to reach one ancestor, called the ``most recent common ancestor''
(MRCA), beginning with $n_{0}-$individuals. 

Most of the models, such as Kingman's approximation of WF or the Moran
models are one-step ones where only $W_{n}^{n-1}$ and $W_{n}^{n}$
are non zero. In these models, the coalescent times $t_{n,n-1}$ are
independent and exponentially distributed. The total time $t_{n_{0},1}$
is the sum of these individual times; therefore, its probability density is 
the convolution of the probability densities of the individual times and
can be computed. However, the computation becomes more tedious for
general models such as the original WF\citep{fu2006a}, where the
jump probabilities from state $n$ are not restricted to state $n-1$
and the independence hypothesis of individual times is no longer valid.
More intricate methods have been developed to tackle the issue of
``multiple merger coalescence'' (for a review, see \citep{tellier2014}).
\begin{figure}[h]
\begin{centering}
\includegraphics[width=1\columnwidth]{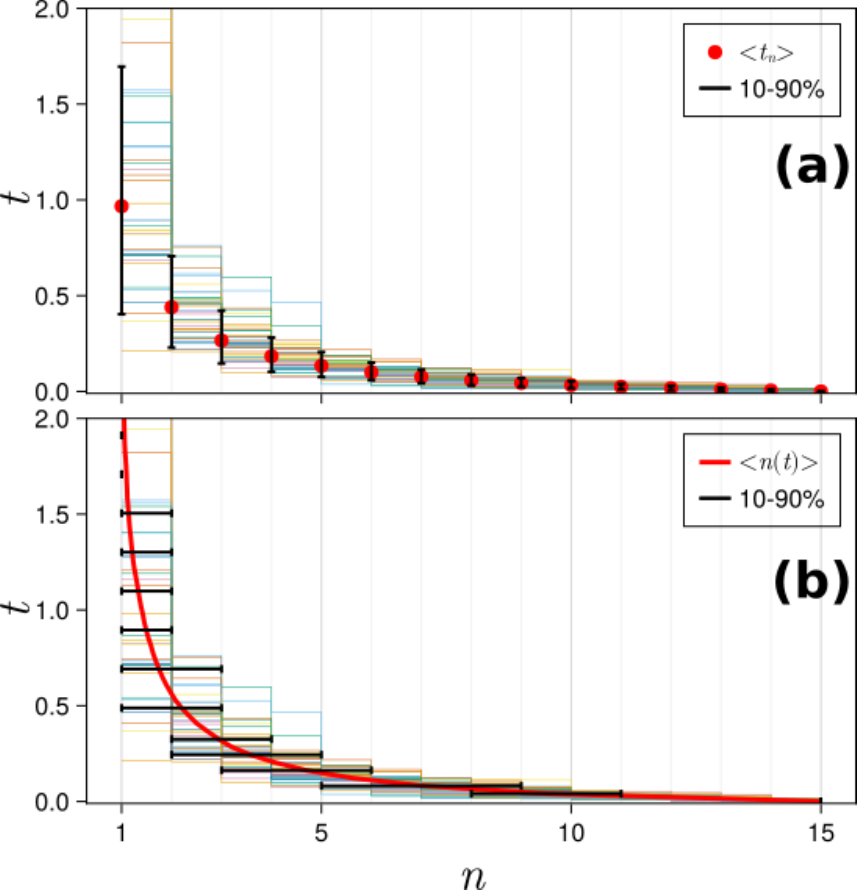}
\par\end{centering}
\caption{Illustration of a Moran coalescent, with initial condition $n_{0}=15$
and population size $N=25$. Background thin lines represent Gillespie
simulations of the random paths $n(t)$. Statistical properties of
the coalescent process can be extracted from these random paths :
(a) vertical slicing: at a given state $n$, the time $t_{n}$ for
each path is recorded; from these times, statistical quantities such
as $\left\langle t_{n}\right\rangle $, the average value of time
to reach state $n$, are computed (here over 1000 paths); vertical
bars denote the 10\%-90\% confidence interval. (b) Horizontal slicing
: at a given time $t$, the state $n(t)$ for each path is recorded.
From these states, statistical quantities such as $\left\langle n(t)\right\rangle $,
the average value of the tree width (number of ancestor) as a function
of time are computed; horizontal bars denote the 10\%-90\% confidence
interval. }\label{fig:Illustration:coalescent}
\end{figure}

The problem of coalescence has been treated in the literature primarily
through the computation of the random variable $t_{n,1}$. I show here
that the problem can be solved generally and in a straightforward
way if instead of computing the probability density $f(t)$ for this
random variable, we instead compute  the probability $P(n,t)$ of observing
$n$ ancestors at time $t$. At first glance, this approach could seem
strange, as we have doubled the number of random variables. However,
as I show below, the evolution of $P(n,t)$, for the most general
models, is governed by a simple Master equation that can be written
as 
\begin{equation}
\Delta_{t}\ket{P(t)}=W\ket{P}\label{eq:master:general}
\end{equation}
where $\ket{P(t)}=(P(1,t),P(2,t),...,P(n_{0},t))^{T}$ is the vector
containing the probabilities for the states, $\Delta_{t}$ is a difference
or differential operator (depending on whether time is continuous
or discrete) and $W$ is a matrix containing the jump probabilities.
What makes the solution particularly simple is the fact that, by construction,
the matrix $W$ is triangular, as the matrix elements $W_{n}^{p}=0$
if $p>n$. In particular, the eigenvalues of the matrix are given
by the diagonal elements. 

We can envision the above stochastic process as an ensemble of (infinite)
random \emph{paths} $n(t)$, where the probabilities and statistical
properties with respect to time or the number of ancestors can be obtained,
both theoretically and numerically, by slicing the curves horizontally
or vertically (figure \ref{fig:Illustration:coalescent}). Probabilities 
$P(n,t)$ contain the most complete information about the system.
For example, the function $P(1,t)$ is the cumulative distribution
function (CDF) of time to reach the MRCA (see below) and its various
moments are simply obtained by (figure \ref{fig:Illustration:coalescent}a)
\begin{eqnarray*}
\left\langle t^{k}\right\rangle  & = & \int_{0}^{\infty}\tau^{k}\frac{dP(1,\tau)}{d\tau}d\tau
\end{eqnarray*}
On the other hand, the moments the random variable $n(t)$ of being
in state $n$ at time $t$ reads
\begin{eqnarray}
\left\langle \left(n(t)\right)^{k}\right\rangle  & = & \sum_{i=1}^{n}i^{k}P(i,t)\label{eq:Moments:def1}
\end{eqnarray}
 and characterize the width of the tree at time $t$ (Figure \ref{fig:Illustration:coalescent}.b). 

This article is organized as follows. In section \ref{sec:Moran-Coalescent},
I show how to solve the master equation (\ref{eq:master:general})
for one-step, continuous models such as the Moran model. Section \ref{sec:Wright=002013Fisher-coalescent.}
is devoted to discrete time models such as the Wright-Fisher one.
The results of sections \ref{sec:Moran-Coalescent} and \ref{subsec:Kingman's-approximation}
are intended to introduce the general methods to solve the mater equation.
These tools are then used in subsection \ref{subsec:The-General-solution.}
to generalize the solution to arbitrary multi-step models. The expressions
in these sections are given in general terms ; however, for comparison
to numerical simulations of the stochastic processes, I use the Moran
and WF models. Section \ref{sec:conclusion-and-discussion} is devoted
to the discussion in which some possible generalizations are considered.
The appendices contain, in a more detailed way, some of the computation
of the article. 

\section{Continuous time, one-step Coalescent}\label{sec:Moran-Coalescent}

\begin{figure}
\begin{centering}
\includegraphics[width=0.75\columnwidth]{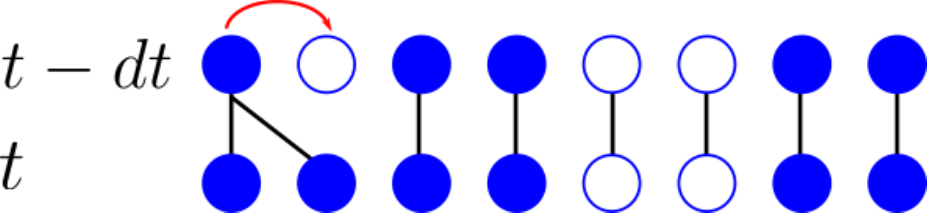}
\par\end{centering}
\caption{Scheme of the Moran model.}\label{fig:Scheme:Moran}

\end{figure}

Consider a one-step, continuous time coalescent process where the
jump probability $W(n\rightarrow n-1)$ is specified. A prime example
of such a process is the continuous time Moran model, which is one
of the two main models of evolutionary biology\citep{moran1962,houchmandzadeh2010}.
The relation between the Moran model and the WF one is discussed in
depth by Alexandre et al.\citep{alexandre2025}. In this model, birth--and-death
events occur simultaneously with a rate $\mu$, where the progeny
of one individual replaces another in the community and the size
of the community remains constant (figure \ref{fig:Scheme:Moran}).

\subsection{Solving the Master Equation.}

For this process, the jump probability is given by\citep{wakeley2009}:
\[
W(n\rightarrow n-1)=W_{n}=\frac{\mu}{N^{2}}n(n-1)
\]
Let us note $P(n,t)$ the probability of observing $n$ ancestors
at time $t$. These probabilities obey the Master Equation in (backward)
time \citep{gardiner2004a}
\begin{eqnarray}
\frac{dP(n_{0},t)}{dt} & = & -W_{n_{0}}P(n_{0},t)\label{eq:ME:moran:1}\\
\frac{dP(n,t)}{dt} & = & W_{n+1}P(n+1,t)\label{eq:ME:moran:2}\\
 &  & -W_{n}P(n,t)\,\,\,\,\,\,\,\,1\le n<n_{0}\nonumber 
\end{eqnarray}
that we can write, in matrix notation, 
\begin{equation}
\frac{d}{dt}\ket{P(t)}=W\ket{P(t)}\label{eq:ME:moran:matricial}
\end{equation}
Where $W$ is the triangular (here bi-diagonal) jump probability matrix
and $n_{0}$ is the number of individuals at time $t=0$. As $W_{1}=0$,
we do not have to distinguish the state $n=1$ in the bulk equation
(\ref{eq:ME:moran:2}). 

As the eigenvalues of $W$ are its diagonal elements, the solution
must be of the form
\begin{equation}
P(n,t)=\sum_{k=n}^{n_{0}}A_{k}^{n}e^{-W_{k}t}\label{eq:moran:direct:t;n}
\end{equation}
that can be written in matrix form as 
\begin{equation}
\ket{P(t)}=A\ket{u(t)}\label{eq:P:vectorial}
\end{equation}
where $\ket{P(t)}=(P(1,t),\ldots P(n_{0},t))^{T}$, the amplitude
$A$ is the triangular matrix with elements $A_{k}^{n}$ and $\ket{u(t)}=(1,e^{-W_{2}t},\ldots,e^{-W_{n_{0}}t})^{T}$. 

For $n=n_{0}$, equation \ref{eq:ME:moran:1} is a simple differential
equation with solution
\begin{equation}
P(n_{0},t)=e^{-W_{n_{0}}t}\label{eq:moran:direct:t:n0}
\end{equation}
and therefore $A_{n_{0}}^{n_{0}}=1.$ 

\begin{figure}
\begin{centering}
\includegraphics[width=0.66\columnwidth]{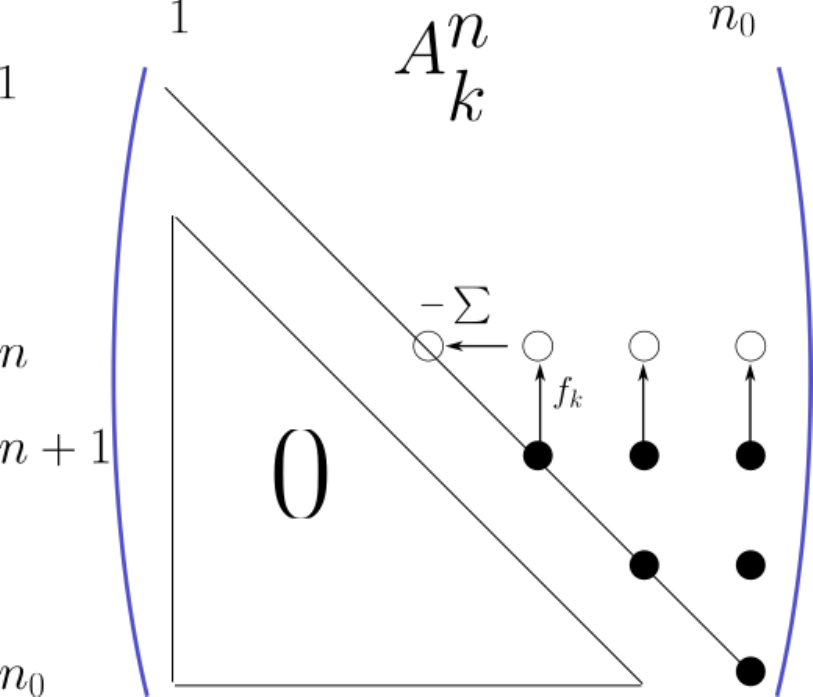}
\par\end{centering}
\caption{Graphical illustration of the construction of the triangular amplitude
matrix $A$. Beginning with line $n_{0}$ where $A_{n_{0}}^{n_{0}}=1$,
each row $n$ is obtained from the row below by multiplying the element
by the factor $f_{k}=W_{n+1}/(W_{n}-W_{k})$. The diagonal element
is obtained by summing the non-{}-diagonal elements of each row and
inverting the sign. The matrix has the following properties : each
column (except the first one) and each row (except the last one) sums
up to zero, and $A_{1}^{1}=1$.  }\label{fig:Graphical-Matrix}

\end{figure}

To obtain the coefficients $A_{k}^{n}$ for $1\le n<n_{0}$, we plug
the solution (\ref{eq:moran:direct:t;n}) into the bulk relation (\ref{eq:ME:moran:2}).
Equating the terms in $\exp(-W_{k}t)$, one finds that
\begin{equation}
A_{k}^{n}=\frac{W_{n+1}}{W_{n}-W_{k}}A_{k}^{n+1}\,\,\,\,\,\,k>n\label{eq:moran:Ank}
\end{equation}
Finally, to compute the remaining coefficient $A_{n}^{n}$, the condition
$P(n,0)=0$ is used :
\begin{equation}
A_{n}^{n}=-\sum_{k=n+1}^{n_{0}}A_{k}^{n}\label{eq:moran:Ann}
\end{equation}
Note that as $P(1,t)\rightarrow1$ when $t\rightarrow\infty$, we
must have $A_{1}^{1}=1.$ Figure \ref{fig:Graphical-Matrix} shows
graphically the construction of the amplitude matrix $A$.

\begin{figure}
\begin{centering}
\includegraphics[width=1\columnwidth]{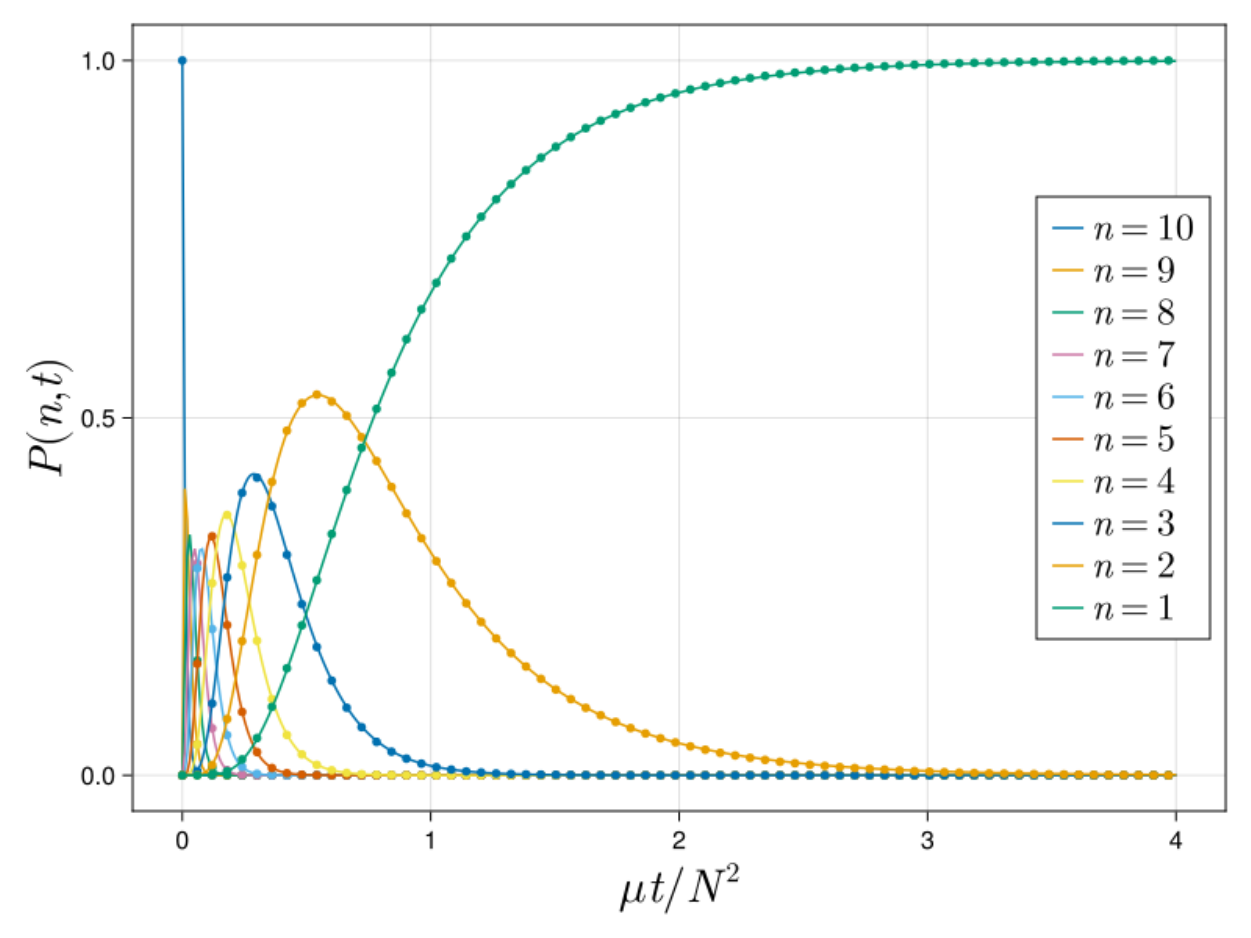}
\par\end{centering}
\caption{Comparison of the theoretical expression (\ref{eq:moran:direct:t;n})
for the probabilities $P(n,t)$ to numerical simulations of the stochastic
process using the Gillespie algorithm, for $N=50$ and $n_{0}=10$.
The numerical probabilities are computed from $10^{5}$ random paths.
}\label{fig:Pnt:theory:sim}
\end{figure}

Figure \ref{fig:Pnt:theory:sim} shows the accuracy of the solution
(\ref{eq:moran:direct:t;n}) compared to the Gillespie stochastic numerical
simulations.

The triangular system (\ref{eq:ME:moran:matricial}) can be solved
directly, without using knowledge of the eigenvalues (see appendix
\ref{subsec:Direct-solution-Moran}). 

\subsection{Time to reach state $n$}

\begin{figure}
\begin{centering}
\includegraphics[width=1\columnwidth]{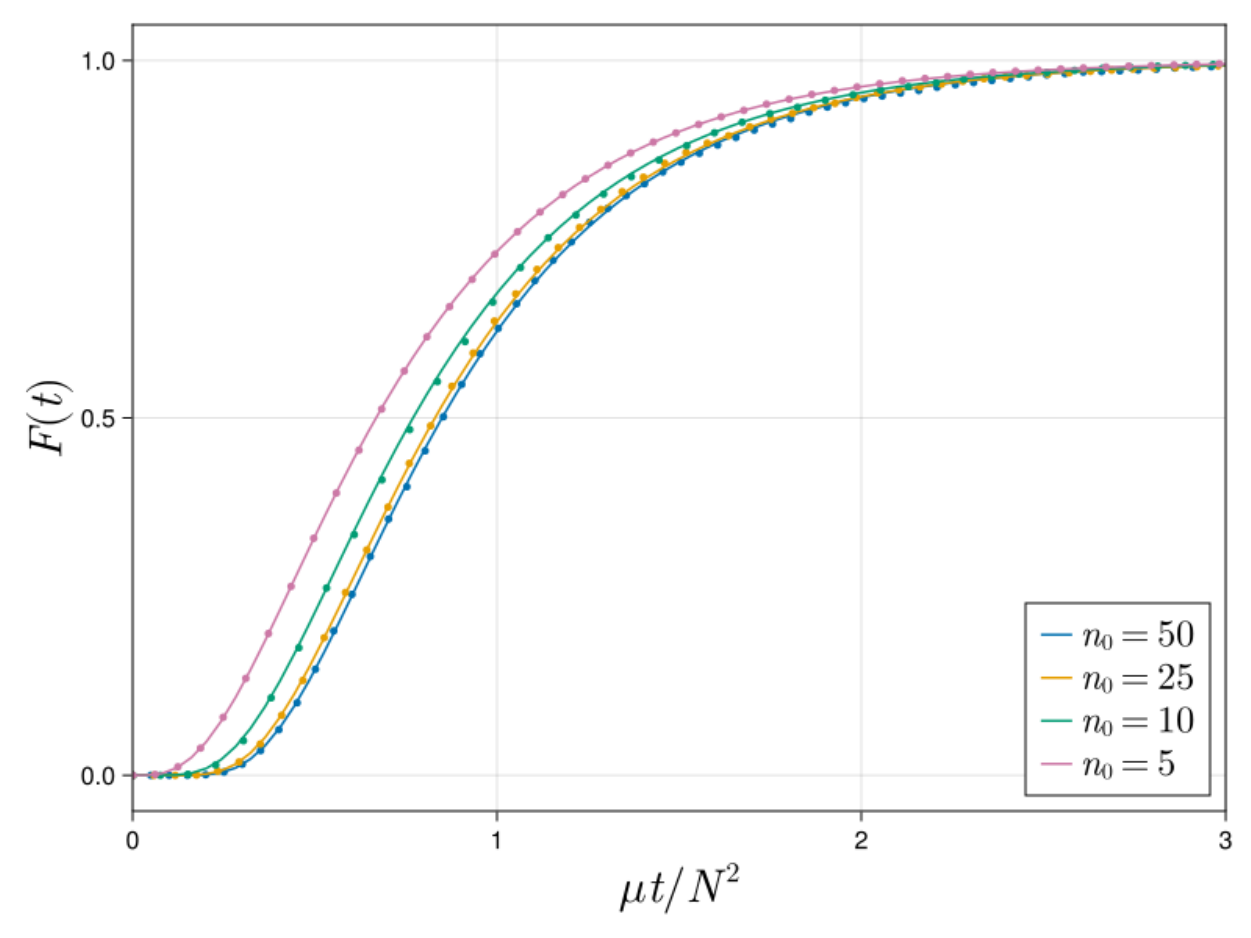}
\par\end{centering}
\caption{The CDF of the coalescent time $t_{1}$(relation \ref{eq:CDF}) compared
to stochastic numerical simulation, for various combination of $n_{0}$
and $N$ . For the numerical simulations, $M$ (here $10^{5}$) random
paths are generated by the Gillespie algorithm and the absorption
time $t_{1}$ for each path is recorded. }\label{fig:CDF:t1}
\end{figure}

Once $P(n,t)$ is known, the time to reach state $n$ can be directly
obtained. Indeed, the function 
\begin{equation}
F_{n}(t)=\sum_{k=1}^{n}P(k,t)\,\,\,\,\,\,n<n_{0}\label{eq:CDF:n}
\end{equation}
is the cumulative distribution function (CDF) of time to reach the state
$n$, beginning at the state $n_{0}$ at $t=0$ (see the appendix \ref{sec:Cumulative-distribution-function}).
In particular, the CDF of coalescent time to reach the MRCA is 
\begin{eqnarray}
F_{1}(t)=P(1,t) & = & \sum_{k=1}^{n_{0}}A_{k}^{1}\exp\left(-W_{k}t\right)\label{eq:CDF}\\
 & = & 1+\sum_{k=2}^{n_{0}}A_{k}^{1}\exp\left(-W_{k}t\right)\label{eq:CDF2}
\end{eqnarray}
as $W_{1}=0$ and $A_{1}^{1}=1.$ Figure \ref{fig:CDF:t1} shows the
accuracy of this expression compared to stochastic numerical simulations.

The moments of the coalescence time can be extracted from the expression
(\ref{eq:CDF2}):

\begin{eqnarray}
\left\langle t^{p}\right\rangle  & = & \int_{0}^{\infty}t^{p}\frac{d}{dt}F_{1}(t)dt\label{eq:moement:t:p}\\
 & = & -p!\sum_{k=2}^{n_{0}}\frac{A_{k}^{1}}{\left(W_{k}\right)^{p}}
\end{eqnarray}
 Analog expressions (for Kingman's coalescent) are discussed in
the literature \citep{mohle2015,pogany2017} with more elaborate methods.
Note that expression (\ref{eq:CDF:n}) is valid only for one-step
processes, while expression (\ref{eq:CDF}) for the time to reach
MRCA is always valid. 

For a \emph{one step process} such as the Moran one, expression (\ref{eq:CDF})
could have been obtained directly \citep{takahata1985b} : the time
$t$ to reach state $k=1$ is the sum of individual waiting times
$t_{k}$ ($1<k\le n_{0})$ on each individual state)
\[
t=\sum_{k=2}^{n_{0}}t_{k}
\]
where $t_{n}$ are \emph{independent} random variables with distribution
probability density 
\[
f_{k}(t)=W_{k}\exp(-W_{k}t).
\]
$f(t)$, the probability distribution of $t$, is therefore the convolution
product of the functions $f_{k}()$, which in the Laplace domain (see
appendix \ref{sec:Computing-the-Amplitude}), reduces to their normal
product:
\[
\hat{f}(s)=\prod_{k=2}^{n_{0}}\frac{W_{k}}{s+W_{k}}
\]
The CDF is in the Laplace domain $\hat{F}(s)=\hat{f}(s)/s$. Its
decomposition into simple fractions and then taking the inverse Laplace
transform leads, without surprise, to the expression (\ref{eq:CDF}). 

Another quantity of interest is the random variable \emph{total length
of all branches} $t_{T}$, which for one-step processes is defined
as 
\begin{equation}
t_{T}=\sum_{k=2}^{n_{0}}kt_{k}\label{eq:Moran:total_branch_length}
\end{equation}
where $t_{k}$ is the random transition time from state $k$ to $k-1$.
The CDF of $t_{T}$ can be recovered by the same computations as above,
by defining an associated stochastic process where the transition
rates are defined as 
\begin{equation}
W'_{n}=W_{n}/n\label{eq:Moran:adjoined}
\end{equation}
The previous computations can be exactly extended to this new process
and, in particular, $P'(1,t)$ is the CDF of the random variable $t_{T}$.
This approach is confirmed by numerical simulations. I will discuss
it in more depth in the next section. 

\subsection{Moments of random variable $n(t)$}

\begin{figure}
\begin{centering}
\includegraphics[width=1\columnwidth]{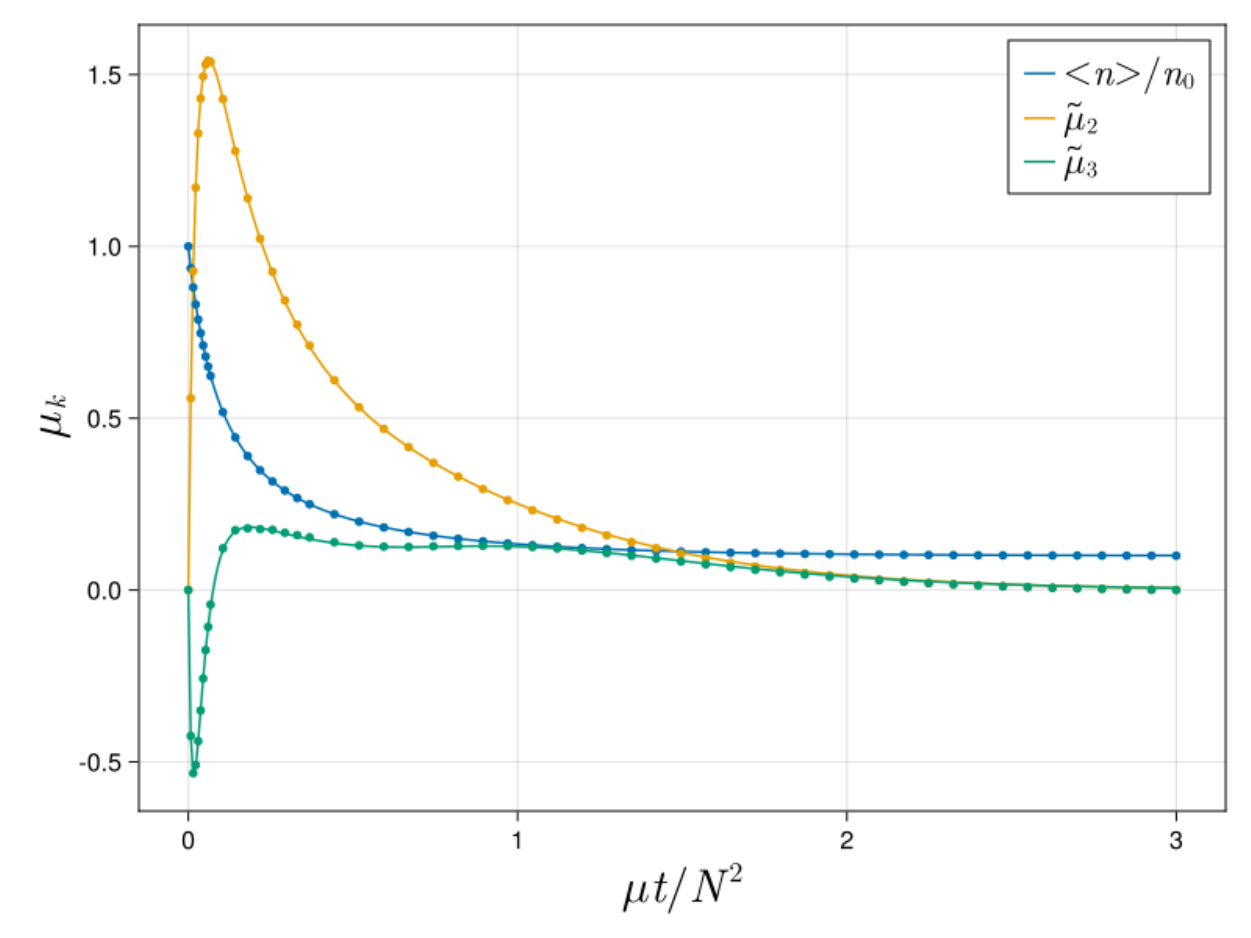}
\par\end{centering}
\caption{The normalized mean $\left\langle n(t)\right\rangle /n_{0}$, the
variance and the third centered moment of $n(t)$ as a function of
time, for $n_{0}=10$. Solid lines are theoretical values computed
from expression (\ref{eq:moment:nk}), dots represents values extracted
from stochastic Gillespie simulations with $10^{5}$ paths.}\label{fig:Moran:nk}

\end{figure}

Knowing the probabilities $P(n,t)$, the moments of the width of the
tree $n(t)$, are easily~computed : 
\begin{equation}
\left\langle \left(n(t)\right)^{p}\right\rangle =\left\langle n^{p}|P(t)\right\rangle =\bra{n^{p}}A\ket{u(t)}\label{eq:moment:nk}
\end{equation}
where $\bra{n^{p}}$ is the linear form (row vector) $(1^{p},2^{p},...,n_{0}^{p})$.
The first three \emph{centered} moments are shown in figure \ref{fig:Moran:nk}.

\section{Discrete time coalescent.}\label{sec:Wright=002013Fisher-coalescent.}

Many models of evolutionary biology are stated in discrete time (generations).
In these models, one specifies the jump probability 
\begin{equation}
W(n\rightarrow k)=W_{n}^{k}\label{eq:WF:Wnk:general}
\end{equation}
of having $k$ ancestor in generation $t-1$, given that there are
$n$ individuals in generation $t$. The master equation in (backward)
discrete time $t$ capturing the number of parents $n$ is therefore the following:
\begin{equation}
P(n,t+1)=W_{n}^{n}P(n,t)+\sum_{j=n+1}^{n_{0}}W_{j}^{n}P(j,t)\label{eq:ME:WF}
\end{equation}
that we can write, in matrix notation, 
\begin{equation}
\ket{P(t+1)}=W\ket{P(t)}\label{eq:WF:matricial}
\end{equation}
where $W$ is the \emph{upper triangular} jump probabilities (relation
\ref{eq:WF:Wnk}). 

The most studied such model is the Wright--Fisher (WF) model of evolution\citep{fisher1999}:
The community is made up of $N$ individuals. In each generation, each
individual produces $p$ progeny and exactly $N$ individuals are
selected among all the progeny to form the next generation. If $p$
is large compared to $N$, the model is simplified : each individual
in generation $t$ picks its parent at random in generation $t-1$.
Now consider $n$ individual in generation $t$, partition them into
exactly $k$ subset, and for each subset, choose one distinct parent
in generation $t-1$ at random (figure \ref{fig:WF:scheme}). The
probability for $n$ individuals in generation $t$ to have exactly
$k$ parents in generation $t-1$ is \citep{hein2005} :
\begin{figure}
\includegraphics[width=0.9\columnwidth]{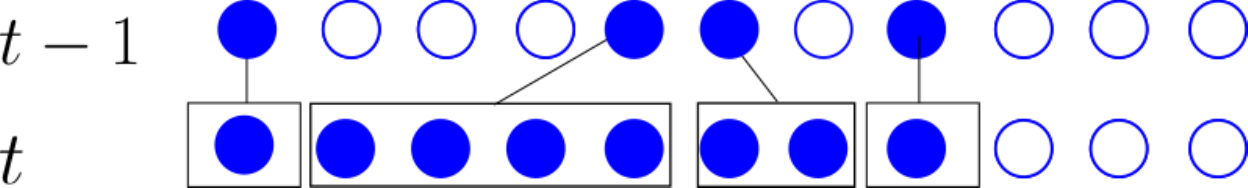}

\caption{The WF coalescent scheme. $n$ individuals are partitioned into $k$
distinct subset ($S(n,k)$ possible configurations) and each subset
chooses a distinct parent ($(N)_{k}$ possible configuration). The
total number of configuration to choose the parents is $N^{n}$, hence
the jump probability of expression (\ref{eq:WF:Wnk})}\label{fig:WF:scheme}

\end{figure}
\begin{equation}
W(n\rightarrow k)=W_{n}^{k}=S(n,k)\frac{(N)_{k}}{N^{n}}\label{eq:WF:Wnk}
\end{equation}
where $S(n,k)$ is the Stirling number of second kind \citep[Section 26.8]{zotero-item-698}that
gives the number of partitions and $(x)_{r}$ is the falling factorial.
\[
(x)_{r}=x(x-1)...(x-r+1)
\]

In the following, I present the solution of this master equation,
first for a one step approximation such as the Kingman's one, and then
generalize the result to the complete equation. 

\subsection{One step approximation.}\label{subsec:Kingman's-approximation}

For large populations $N\gg1$, Kingman's approximation\citep{kingman1982}
consists of keeping only $O(1/N)$ terms in the jump probabilities.
The only terms remaining are 
\begin{eqnarray*}
W_{n}^{n} & = & 1-\frac{1}{2N}n(n-1)+O(1/N^{2})\\
W_{n}^{n-1} & = & \frac{1}{2N}n(n-1)+O(1/N^{2})
\end{eqnarray*}
and in effect, the system reduces to a one-step model similar to the
Moran's one. The master equation in this framework is therefore
\begin{eqnarray}
P(n_{0},t+1) & = & W_{n_{0}}^{n_{0}}P(n_{0},t)\label{eq:ME:kingman}\\
P(n,t+1) & = & W_{n}^{n}P(n,t)+W_{n+1}^{n}P(n+1,t)\label{eq:ME:kingman:bulk}\\
 &  & 1\le n<n_{0}\nonumber 
\end{eqnarray}
The above equations have the same form as the Moran master equation
and can be solved by similar methods. In particular, if we use the
Kingman's second approximation by using continuous times and setting
\[
P(n,t+1)-P(n,t)\approx\frac{\partial P(n,t)}{\partial t}
\]
the problem is mapped exactly to the one treated in the preceding
section. However, this approximation is not necessary. Using again
the triangular nature of the system and the fact that eigenvalues
of $W$ are its diagonal elements, the solution must be of the form
\begin{eqnarray}
P(n_{0},t) & = & \left(W_{n_{0}}^{n_{0}}\right)^{t}\label{eq:WFK:sol:n0:t}\\
P(n,t) & = & \sum_{k=n}^{n_{0}}A_{k}^{n}\left(W_{k}^{k}\right)^{t}\,\,\,n<n_{0}\label{eq:WFK:sol:n:t}
\end{eqnarray}
The coefficients $A_{k}^{n}$ (for $n<n_{0}$ and $k>n$) are found
by integrating expression (\ref{eq:WFK:sol:n:t}) into relation (\ref{eq:ME:kingman:bulk})
and grouping terms in $(W_{k}^{k})^{t}$ : 
\begin{eqnarray}
A_{n_{0}}^{n0} & = & 1\label{eq:WFK:Ank}\\
A_{k}^{n} & = & \frac{W_{n+1}^{n}}{W_{k}^{k}-W_{n}^{n}}A_{k}^{n+1}\,\,\,\,\,\,\,\,k>n
\end{eqnarray}
Using the fact that $P(n,0)=0$, the coefficient $A_{n}^{n}$ is found
\begin{equation}
A_{n}^{n}=-\sum_{k=n+1}^{n_{0}}A_{k}^{n}\label{eq:WFK:Kingman:Ann}
\end{equation}
If we recall $W_{j}^{j}=1-W_{j}^{j-1}$, we see that expressions
(\ref{eq:WFK:Ank},\ref{eq:WFK:Kingman:Ann}) are formally the same
as those obtained for the Moran model. The appendix \ref{subsec:Direct:solution:WF}
shows how the above solution can be obtained directly and without
using the prior knowledge of the eigenvalues. 

\subsection{The General solution. }\label{subsec:The-General-solution.}

There is no difference between the solution in the one step approximation
framework (such as Kingman's ) and the general one. In both cases,
we only need to use the triangular nature of the system : the ``bi--diagonal''
nature of the one step approximation does not bring any particular
simplification. The solution of the master equation (\ref{eq:ME:WF})
must still be of the form given by equations (\ref{eq:WFK:sol:n0:t},\ref{eq:WFK:sol:n:t})
. To compute the amplitudes, we plug these relations into the master
equation: 
\begin{eqnarray*}
\sum_{k=n+1}^{n_{0}}\left(W_{k}^{k}-W_{n}^{n}\right)A_{k}^{n}\left(W_{k}^{k}\right)^{t} & = & \sum_{j=n+1}^{n_{0}}W_{j}^{n}\sum_{k=j}^{n_{0}}A_{k}^{j}\left(W_{k}^{k}\right)^{t}\\
 & = & \sum_{k=n+1}^{n_{0}}\left(W_{k}^{k}\right)^{t}\sum_{j=n+1}^{k}W_{j}^{n}A_{k}^{j}
\end{eqnarray*}
 where the second line is obtained by permuting the summation index.
The amplitudes $A_{k}^{n}$ are again found by grouping the terms
in $\left(W_{k}^{k}\right)^{t}$:

\begin{eqnarray}
A_{n_{0}}^{n_{0}} & = & 1\label{eq:WF:rec1}\\
A_{k}^{n} & = & \frac{1}{W_{k}^{k}-W_{n}^{n}}\sum_{j=n+1}^{k}W_{j}^{n}A_{k}^{j}\,\,\,\,\,\,\,\,k>n\label{eq:WF:rec2}\\
A_{n}^{n} & = & -\sum_{k=n+1}^{n_{0}}A_{k}^{n}\,\,\,\,\,\,\,\,k>n\label{eq:WF:rec3}
\end{eqnarray}
where (\ref{eq:WF:rec3}) is obtained from condition $P(n,0)=0$. 

Comparing expressions (\ref{eq:WFK:Ank}) and (\ref{eq:WF:rec2}),
we see that the only difference is that, to calculate $A_{k}^{n}$,
we need all the known coefficients $A_{k}^{j}$ from the lines below
and not only the coefficients from the line $n+1$. This recurrence
naturally disentangles the entanglements introduced by allowing jumps
from state $n$ to any inferior states. It is straightforward to check
that the above expression reduces to the relation (\ref{eq:WFK:Ank})
for the Kingman's approximation where $W_{j}^{n}=0$ if $j>n+1$. 

\begin{figure}
\begin{centering}
\includegraphics[width=1\columnwidth]{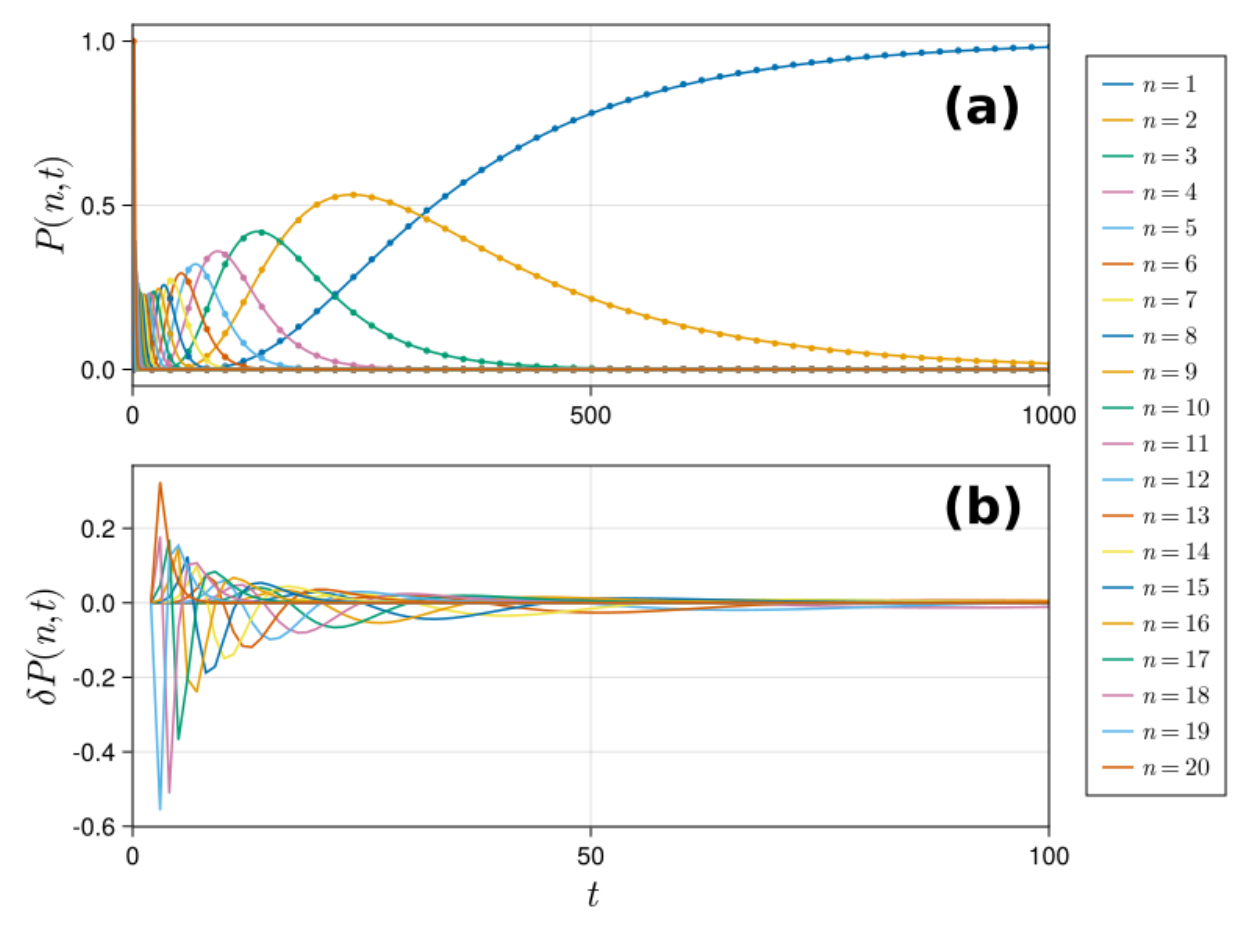}
\par\end{centering}
\caption{Probabilities $P(n,t)$. Two sets of data are generated by a Gillespie
algorithm, one for the stochastic process governed by the Kingman's
approximation and one for the general WF model, and they are compared
to theoretical expressions obtained from the amplitude matrix in each
case. (a) comparison of the general solution (solid lines) to numerical
solutions (circles). The accuracy is similar for the Kingman's approximation
and is not displayed. (b) Comparison of the theoretical probabilities
for the General and Kingman's approximation, where $\delta P=P_{\text{General}}-P_{\text{Ki,gman's}}$.
$N=200$; $n_{0}=20$. }\label{fig:WF:Pnt}
\end{figure}

Figure \ref{fig:WF:Pnt} displays the comparison of the theoretical
probabilities of the general solution and Kingman's approximation
to numerical simulations.

\subsection{Time to reach the MRCA}

The problem of time to reach MRCA for the general WF process has been
treated by Fu\citep{fu2006a}, using a direct approach. Here, I show
that the time to reach the state $n$ can be calculated simply using an
 approach similar to that in the preceding section. Note that for the
general model, contrary to one step models, only the absorption time
$t$ to reach the state $n=1$ (the MRCA) is unambiguously defined. As
in the previous section, the CDF of this random variable is (see appendix
\ref{sec:Cumulative-distribution-function})
\[
F(t)=P(1,t)
\]
which is the probability that the system is in state $n=1$ at time
$t$ (has reached $n$ at time $\tau<t$). The probability of reaching
$n=1$ exactly at $t$ is therefore 
\begin{eqnarray*}
f(0) & = & 0\\
f(t) & = & \Delta F=F(t)-F(t-1)
\end{eqnarray*}
The moments of the random variable $t$ are 
\[
\left\langle t^{p}\right\rangle =\int_{0}^{\infty}t^{p}f(t)dt
\]
and can be given in terms of the polylogarithm function. However, it is 
simpler to use the $Z-$transform (probability generating function),
defined as 
\[
\hat{f}(s)=\sum_{t=0}^{\infty}s^{-t}f(t)
\]
Defining the operator $L=s\,d/ds$, the moments of the random variable
$t$ are obtained by a multiple application of this operator to $\hat{f}(s)$
: 
\[
\left\langle t^{p}\right\rangle =(-1)^{p}\left.L^{p}\hat{f}(s)\right|_{s=1}
\]
The expression for $\hat{f}(s)$ reads: 
\begin{eqnarray*}
\hat{f}(s) & = & 1+\sum_{k=2}^{n_{0}}A_{k}^{1}\frac{s-1}{s-W_{k}^{k}}
\end{eqnarray*}
In particular, the first moments are 
\begin{eqnarray*}
\left\langle t\right\rangle  & = & -\sum_{k=2}^{n_{0}}\frac{A_{k}^{1}}{1-W_{k}^{k}}\\
\left\langle t^{2}\right\rangle  & = & -\sum_{k=2}^{n_{0}}A_{k}^{1}\frac{1+W_{k}^{k}}{\left(1-W_{k}^{k}\right)^{2}}\\
\left\langle t^{3}\right\rangle  & = & -\sum_{k=2}^{n_{0}}A_{k}^{1}\frac{1+4W_{k}^{k}+(W_{k}^{k})^{2}}{\left(1-W_{k}^{k}\right)^{3}}
\end{eqnarray*}
The difference between the general case and the Kingman's approximation
is reflected in the computation of the amplitudes $A_{k}^{1}$. 

\begin{figure}
\begin{centering}
\includegraphics[width=1\columnwidth]{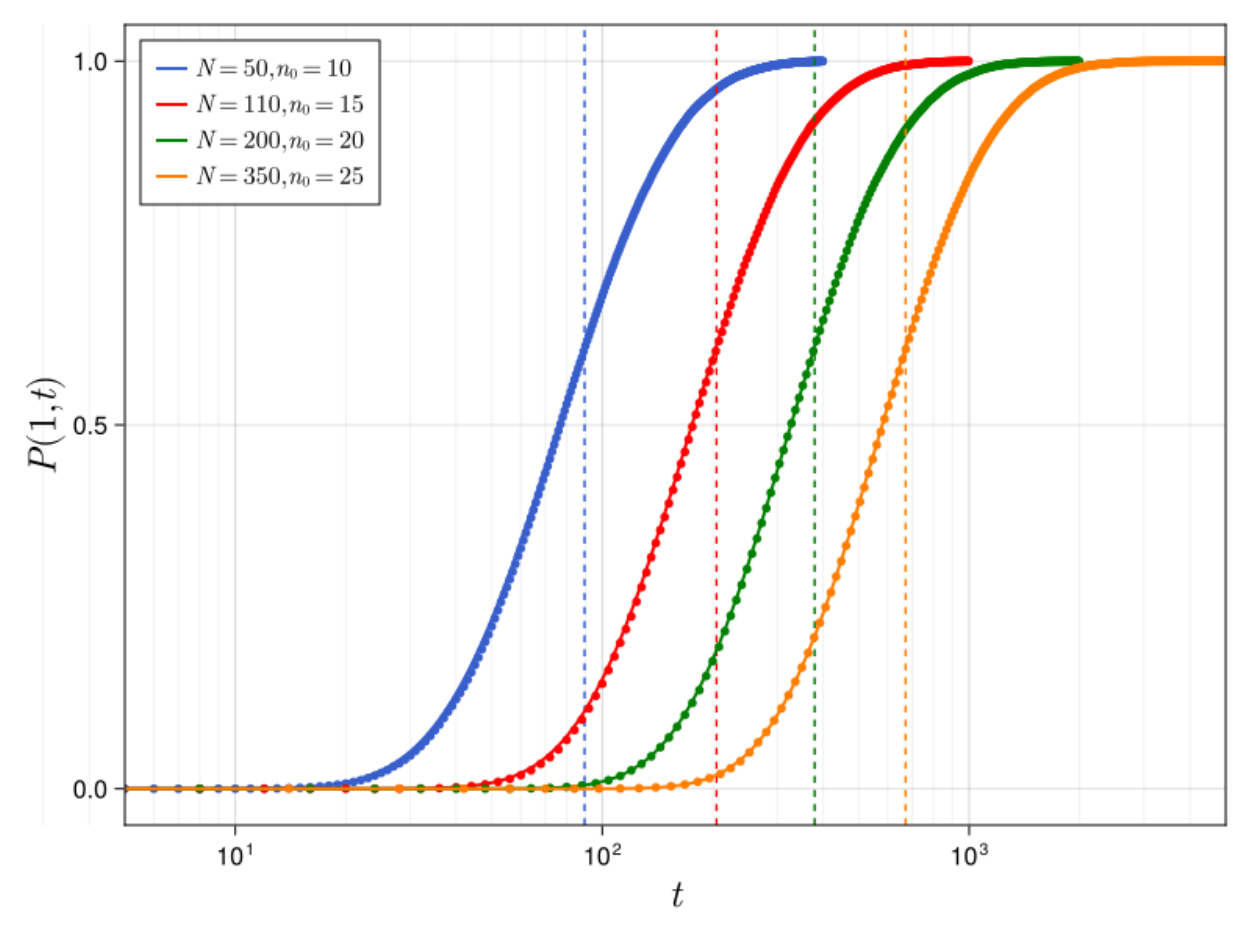}
\par\end{centering}
\caption{The CDF of the random variable $t_{1}$ as a function of time for
various combination of parameters. Solid lines:theoretical ; circles
: numerical simulation. Vertical dashed lines are the mean time $\left\langle t\right\rangle $
for each set of parameters.}\label{fig:WF:F1t}
\end{figure}

Figure \ref{fig:WF:F1t} displays the CDF and the mean of $t$ for
various combinations of $n_{0}$ and $N$. These relations are confirmed
by numerical simulations. Computing numerically $\left\langle t\right\rangle $
for a wide combination of $n_{0}$ and $N$, we note that the difference
in coalescence time $\left\langle t\right\rangle $ between Kingman's
approximation and the general solution is $O(1)$. The Kingman's approximation
is indeed an excellent one, as was already noted by Fu\citep{fu2006a}. 

As mentioned in the previous section, another quantity of interest
is the random variable $t_{T}$, the total length of all branches.
For a one step process, this quantity is unambiguously defined as
$t_{T}=\sum_{j=2}^{n_{0}}jt_{j}$, where $t_{j}$ is the random transition
time from state $j$ to $j-1$, \emph{i.e.} , the time the system remains
in state $j$. For general processes, this definition is no longer valid,
as the system can transition from a state $k$ to an arbitrary state
$j$ ($j<k$). 

\begin{figure}
\begin{centering}
\includegraphics[width=1\columnwidth]{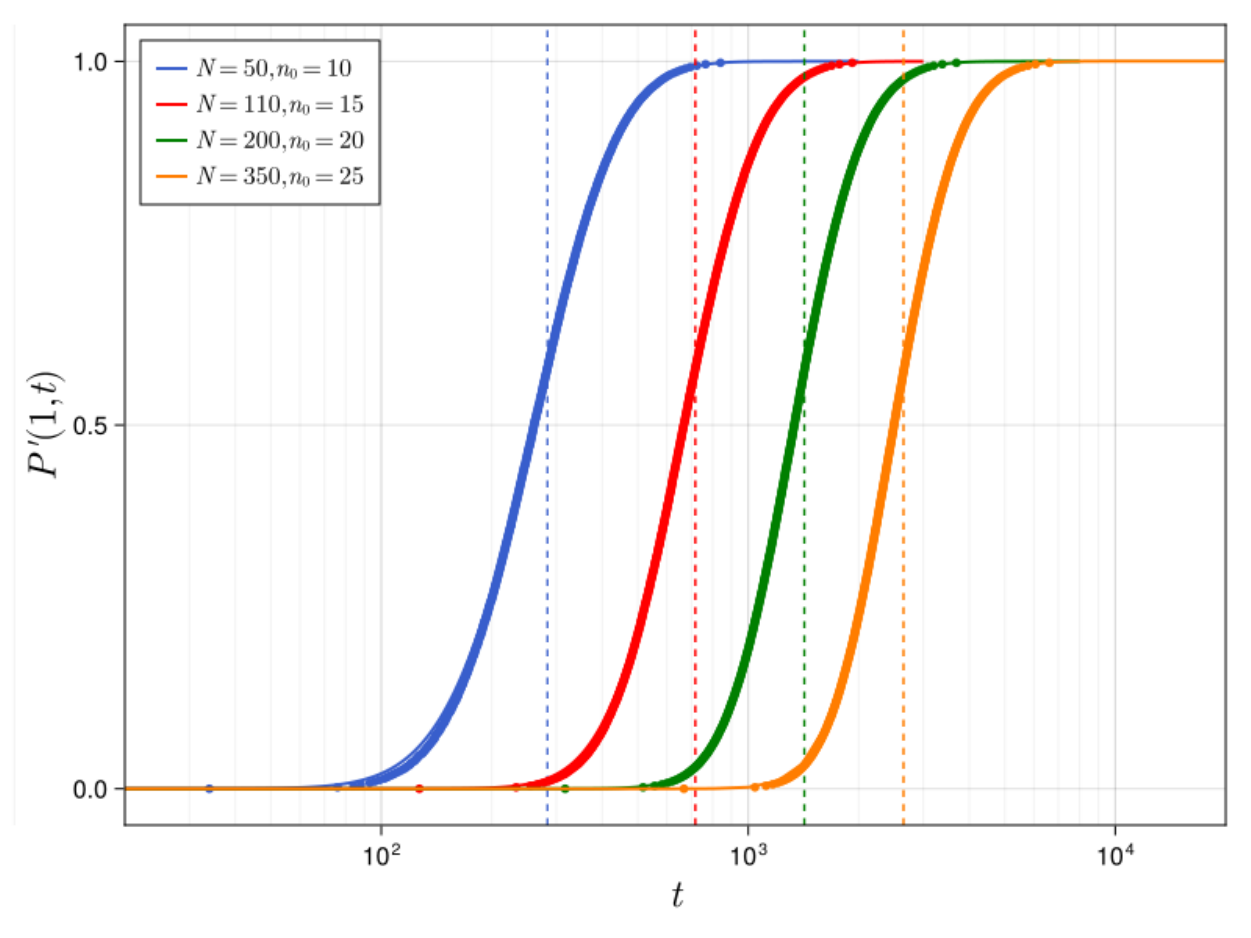}
\par\end{centering}
\caption{The CDF of the random variable $t_{T}$, total branch length, as a
function of time for various combination of parameters. Solid lines:theoretical
; circles : numerical simulation. Vertical dashed lines are the mean
time $\left\langle t_{T}\right\rangle $ for each set of parameters.
Numerical simulation is performed by making $M=10^{5}$ realizations
of the General coalescent process, computing $t_{T}$ for each realization
according to relation (\ref{eq:WF:tT:num}) . Theoretical result is
obtained from the associated stochastic process. }\label{fig:CDF:total:length}

\end{figure}

For one random realization of this process, the system goes through
states ($n_{\alpha_{0}},n_{\alpha_{1}},n_{\alpha_{2}},\ldots,1)$
where $n_{\alpha_{0}}=n_{0}$, and for this realization, the total
time is unambiguously defined as
\begin{equation}
t_{T}=\sum_{\alpha_{i}}\alpha_{i}t_{\alpha_{i}}\label{eq:WF:tT:num}
\end{equation}
Numerically, it is straightforward to compute this time for $M$ realizations
and use these $t_{T,i}$ to deduce the CDF of the random variable
$t_{T}$. The procedure used above for the computation of $t_{MRCA}$
can be used to address this problem with the same simplicity. We only
need to define an associated stochastic process in which 
\begin{eqnarray}
W'{}_{n}^{k} & = & W_{n}^{k}/n\,\,\,\,\,\,\,k\neq n\label{eq:WF:W':nk}\\
W'{}_{n}^{n} & = & 1-\sum_{k\neq n}W'{}_{n}^{k}\label{eq:WF:W':nn}
\end{eqnarray}
For this associated process, $P'(1,t)$ is the CDF of the total branch
length. Indeed, for this process,  the random time that the system stays in the state $n$ is multiplied by $n$, regardless of the states to which it transitions and without modifying the relative probabilities of various transitions . Figure \ref{fig:CDF:total:length}
 numerically confirms the validity of this approach. 

\subsection{Moments of random variable $n(t)$}

From the expression of the probabilities, various moments of the 
width of the tree can be obtained. These moments of are defined
as 
\begin{figure}
\begin{centering}
\includegraphics[width=1\columnwidth]{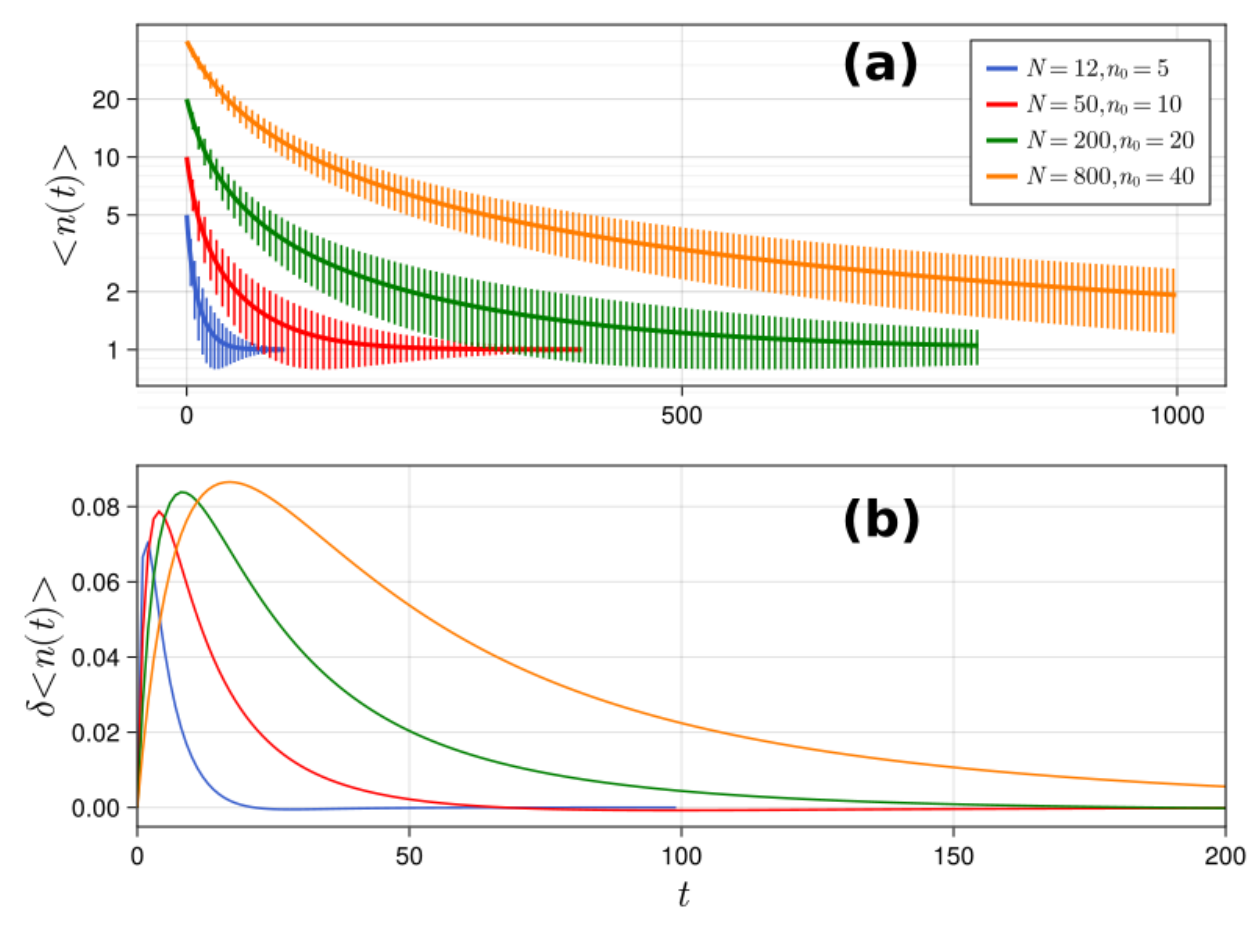}
\par\end{centering}
\caption{Width of the tree $\left\langle n(t)\right\rangle $ for the general
WF model. (a) $\left\langle n(t)\right\rangle $ (solid lines) for
various combinations of $n_{0}$ and $N$ ; vertical bars denote the
standard deviation $\sigma=\sqrt{\left\langle n^{2}\right\rangle -\left\langle n\right\rangle ^{2}}.$
(b) The difference in the computed mean between the general WF model
and the Kingman's approximation. The theoretical expressions are confirmed
by comparison to the results of numerical stochastic equations (not
displayed). }\label{fig:WF_moments_n}
\end{figure}
\[
\left\langle n^{p}(k)\right\rangle =\left\langle n^{p}|P(t)\right\rangle 
\]
where the linear form $\bra{n^{p}}=(1,2^{p},...,n_{0}^{p})$. Figure
\ref{fig:WF_moments_n} displays the mean and standard deviation
of $n(t)$ for various combinations of $n_{0}$ and $N$. 

\section{conclusion and discussion}\label{sec:conclusion-and-discussion}

In this article, I have shown how the coalescent model, for a community
of $N$ individuals, can be solved simply and generally by shifting
the focus from the random variable $t$, time to reach the MRCA, to
the joined variable $(n,t)$ that encompasses the time and width 
of the coalescent tree. It appears that the master equation for $P(n,t)$
can be solved efficiently by recurrence for the most general case,
due to the peculiar nature of jump rates in the coalescent problem.
The solutions of the preceding sections were given in general terms
and were compared to stochastic simulations of two well studied models,
but can be used for any exchangeable model (reviewed for example in
\citep{mohle2001}). For one step processes such as Moran or Kingman's
approximation of WF, the computation is equivalent in terms of complexity
to a direct computation of time to MRCA. On the other hand, for more
general models, the method I presented in the preceding sections conserves
its simplicity and is solved with the same tools that are used for the one
step models. 

An area where this work may be extended is the structured coalescent
(see \citep[chap. 5]{wakeley2009} for a review). Natural populations
are subdivided into geographically distinct sub--populations with
migrations taking place between them, and one can indeed wonder if
the limit of one large population of size $N$ can realistically capture 
such a structure. A natural extension of this work would be to consider
the probability $P(n_{1},\ldots,n_{D};t)$ describing these populations,
where $D$ is the number of subpopulations and $n_{i}$ the number
of ancestors in deme $i$. It is beyond the scope of this article
to investigate this problem. However, at first glance, the downward
nature of jump probabilities indicates that this may be feasible.
The case for $D=2$ seems to be straightforward, but it remains to
be determined if the general case for arbitrary $D$ can conserve
the simplicity and computability of the preceding sections.

\appendix

\section{Direct solution of the master equation}\label{sec:Computing-the-Amplitude}

\subsection{Continuous model}\label{subsec:Direct-solution-Moran}

The systems governing the continuous master equation can be solved
directly using Laplace transforms. The Laplace transform (with respect
to time) of a function $f(t)$ is defined as \citep[Sect. 1.14]{zotero-item-698}
\begin{equation}
\hat{f}(s)=\int_{0}^{\infty}e^{-st}f(t)dt\label{eq:laplace:transform}
\end{equation}
and has the property of transforming linear differential equations
into algebraic ones:
\[
{\cal L}[f'(t)]=s\hat{f}(s)-f(0)
\]
For the Moran model, taking the Laplace transform of equations (\ref{eq:ME:moran:1},\ref{eq:ME:moran:2}),
produces : 
\begin{eqnarray}
\left(s+W_{n_{0}}\right)\hat{P}(n_{0},s) & = & 1\label{eq:a:Pn0s}\\
\left(s+W_{n}\right)\hat{P}(n,s) & = & W_{n+1}\hat{P}(n+1,s)\label{eq:a:Pns}
\end{eqnarray}
The solution of the above equations, obtained by recurrence, is 
\begin{eqnarray}
\hat{P}(n_{0},s) & = & 1/\left(s+W_{n_{0}}\right)\label{eq:Pn0s}\\
\hat{P}(n,s) & = & \prod_{k=n+1}^{n_{0}}W_{k}/\prod_{k=n}^{n_{0}}\left(s+W_{k}\right)\label{eq:Pns}
\end{eqnarray}
The solution (\ref{eq:Pns}) can be decomposed into the sum of simple
fractions 
\begin{equation}
\hat{P}(n,s)=\sum_{k=n}^{n_{0}}\frac{A_{k}^{n}}{s+W_{k}}\label{eq:Pns:solution}
\end{equation}
where 
\begin{eqnarray*}
A_{n_{0}}^{n_{0}} & = & 1\\
A_{k}^{n} & = & \prod_{j=n+1}^{n_{0}}W_{j}/\prod_{j=n,j\ne k}^{n_{0}}\left(W_{j}-W_{k}\right)\\
 &  & \,\,\,\,\,\,n<n_{0}\,\,,\,\,k\ge n
\end{eqnarray*}
We note that as $W_{1}=0,$ $A_{1}^{1}=1$. 

By taking the inverse Laplace transform, we get the explicit solution
for the probabilities:
\begin{equation}
P(n,t)=\sum_{k=n}^{n_{0}}A_{k}^{n}\exp\left(-W_{k}t\right)\label{eq:Pnt:final}
\end{equation}

One of the properties of the amplitude matrix for a one step process
is 
\begin{eqnarray}
-\sum_{k=2}^{n_{0}}\frac{A_{k}^{1}}{W_{k}} & = & \sum_{k=2}^{n_{0}}\frac{1}{W_{k}}\label{eq:sum:A1k:Wk}
\end{eqnarray}

\subsection{Discrete model. }\label{subsec:Direct:solution:WF}

The systems governing the discrete master equation can be solved by
a similar method as above, using the discrete Laplace transforms (Z-transform).
The Z-transform of the discrete function $f(t)$ is defined as 
\[
Z[f(t)]=\hat{f}(s)=\sum_{t=0}^{\infty}s^{-t}f(t)
\]
And its properties are similar to the continuous Laplace transform.
In particular, 
\begin{eqnarray*}
Z[f(t+1)] & = & s\hat{f}(s)-sf(0)\\
Z[\lambda^{t}] & = & \frac{s}{s-\lambda}
\end{eqnarray*}
Applying the $Z-$transform to the relation (\ref{eq:ME:kingman})
and using the condition $P(n_{0},0)=1$, we obtain the following:
\[
\hat{P}(n_{0},s)=\frac{s}{s-W_{n_{0}}^{n_{0}}}
\]
by inverting the transform, we have $P(n_{0},t)=\left(W_{n_{0}}^{n_{0}}\right)^{t}$.
 Applying the $Z-$transform to the relation (\ref{eq:ME:kingman:bulk})
and using the condition $P(n,0)=0$, we obtain the following:
\begin{eqnarray}
\hat{P}(n,s) & = & \frac{W_{n+1}^{n}}{s-W_{n}^{n}}\hat{P}(n+1,s)\label{eq:sub:WFK:recurrence}\\
 & = & s\prod_{k=n}^{n_{0}-1}W_{k+1}^{k}\bigg/\prod_{k=n}^{n_{0}}(s-W_{k}^{k})\nonumber 
\end{eqnarray}
and we can use the decomposition into simple fractions to simplify the
relation. However, it is, as in the preceding subsection, simpler to
 directly obtain the recurrence relation between amplitudes. Suppose
that 
\[
\hat{P}(n,s)=\sum_{k=n}^{n_{0}}A_{k}^{n}\frac{s}{s-W_{k}^{k}}
\]
Then, using the recurrence relation (\ref{eq:sub:WFK:recurrence})
and simple decomposition,
\[
\frac{s}{(s-W_{k}^{k})(s-W_{n}^{n})}=\frac{1}{W_{k}^{k}-W_{n}^{n}}\left(\frac{s}{s-W_{k}^{k}}-\frac{s}{s-W_{n}^{n}}\right)
\]
we obtain the recurrence relation between amplitudes: 
\begin{eqnarray*}
A_{k}^{n} & = & \frac{W_{n+1}^{n}}{W_{k}^{k}-W_{n}^{n}}A_{k}^{n+1}\,\,\,\,\,\,\,\,\,k>n\\
A_{n}^{n} & = & -\sum_{k=n+1}^{n_{0}}A_{k}^{n}
\end{eqnarray*}

\section{Cumulative distribution function.}\label{sec:Cumulative-distribution-function}

Consider the (downward) stochastic processes discussed in previous
sections where $P(n,t)$ is the probability of \emph{observing} the
system in state $n$ at time $t$, with initial condition $P(n,0)=\delta_{n_{0}}^{n}$,
where $\delta$ is the Kronecker symbol. The probability that the
system is in a state $m>n$ (has not yet reached the state $n$) is
\[
Q_{n}(t)=\sum_{k=n+1}^{n_{0}}P(k,t)
\]
and therefore the probability density $f_{n}(t)$ of \emph{reaching}
the state $n$ at $t$ is \citep{gardiner2004a}
\begin{equation}
f_{n}(t)=-\frac{d}{dt}Q_{n}(t)=\sum_{k=1}^{n}\frac{d}{dt}P(k,t)\label{eq:proba:density}
\end{equation}
where the summation boundary has been changed using the relation $\sum_{k=1}^{n_{0}}P(n,t)=1$.
Therefore, the cumulative distribution function of the time to reach state
$n$ is 
\[
F_{n}(t)=\int_{0}^{t}f_{n}(\tau)d\tau=\sum_{k=1}^{n}P(k,t)
\]
The moments of coalescent time $\left\langle t_{1}^{p}\right\rangle $
to reach the MRCA are obtained from the successive derivation of the
Laplace transform of the probability density function : from the definition
of the Laplace transform, we have 
\[
\hat{f}_{1}^{(p)}(s)=(-1)^{p}\int_{0}^{\infty}t^{p}e^{-st}f_{1}(t)dt
\]
and therefore
\[
\left\langle t^{p}\right\rangle =(-1)^{p}\hat{f}_{1}^{(p)}(0)
\]
As we know the explicit expression 
\begin{eqnarray*}
\hat{f}_{1}(s) & = & s\hat{P}(1,s)\\
 & = & 1+\sum_{k=2}^{n_{0}}A_{k}^{1}\frac{s}{s+W_{k}}=-\sum_{k=2}^{n_{0}}A_{k}^{1}\frac{W_{k}}{s+W_{k}}
\end{eqnarray*}
we have 
\[
\hat{f}_{1}^{(p)}(s)=(-1)^{p+1}p!\sum_{k=2}^{n_{0}}A_{k}^{1}\frac{W_{k}}{\left(s+W_{k}\right)^{p+1}}
\]
Evaluating this expression at $s=0$ leads to equation (\ref{eq:moement:t:p}).

For discrete processes, the probability of reaching the state $n$ in exactly
time $t>0$ is 
\[
f_{n}(t)=P(n,t)-P(n,t-1)
\]
Note that for a discrete process, the process of defining $f_{n}$ is
ambiguous and we could have used $P(n,t+1)-P(n,t)$. This definition
introduces a shift of one generation.

\section{Numerical computations.}\label{sec:Numerical-computations.}

There are general numerical packages, such as the one published by
Vaughan and Drummond \citep{vaughan2013} for stochastic processes
that can simulate genealogies. In this article, I have written simple
codes that directly implement the Gillespie algorithm\citep{gillespie1977a}
: When the system is in state $n$, the time to the next transition
follows an exponential (for continuous times) or geometric (for discrete
times) distribution with rate $W=\sum_{k\ne n}W_{n}^{k}$ and is drawn
from the corresponding function. The nature of the transition is obtained
from the normalized rates $W_{n}^{k}/W$ and the new state of the
system is updated accordingly. The procedure continues until the adsorbing
state $n=1$ is reached, and a table containing states and transition
times is returned, corresponding to one stochastic path. $M$ (usually
$10^{3}-10^{5}$) random paths are generated to compute the probabilities
$P(n,t)$. 

All computations and numerical simulations were performed with
the high level language ``Julia''\citep{bezanson2017julia}.

\bibliographystyle{unsrt}
\bibliography{coalescent}

\end{document}